\documentclass[aps,pra,showpacs,twocolumn,groupedaddress]{revtex4}
\usepackage{graphics}
\usepackage{graphicx}
\newcommand \beq{\begin{eqnarray}}
\newcommand \eeq{\end{eqnarray}}
\newcommand \bea{\begin{eqnarray}}
\newcommand \eea{\end{eqnarray}}
\def\simge{\mathrel{%
       \rlap{\raise 0.511ex \hbox{$>$}}{\lower 0.511ex \hbox{$\sim$}}}}
\def\simle{\mathrel{
       \rlap{\raise 0.511ex \hbox{$<$}}{\lower 0.511ex \hbox{$\sim$}}}}

\begin{document}

    \title{The Kosterlitz-Thouless-Berezinskii transition of homogeneous and
trapped Bose gases in two dimensions}

\author{Markus Holzmann,$^a$ Gordon Baym,$^b$ Jean-Paul Blaizot,$^{c,d}$ and
Franck Lalo{\"e}$^e$
\\
$^a$LPTMC, UMR 7600 of CNRS, Universit\'e Pierre et Marie Curie, 4 Place Jussieu,
75005 Paris, France.
\\
$^b$Department of Physics, University of Illinois, 1110 W. Green St.,
Urbana, IL 61801
\\
$^c$ECT*, 38050 - Villazzano (Trento), Italy
\\
$^d$Service de Physique Th\'eorique, CE-Saclay, Orme des Merisiers,
   91191 Gif-sur-Yvette, France
   \\
$^e$LKB, Ecole Normale Sup\'erieure, 24 rue Lhomond, 75005 Paris,
   France}

\date{\today}

\begin{abstract}

    We derive the scaling structure of the Kosterlitz-Thouless-Berezinskii
(KTB) transition temperature of a homogeneous Bose gas in two dimensions
within diagrammatic perturbation theory.  Approaching the system from above
the transition, we calculate the critical temperature, $T_{KT}$, and show how
the superfluid mass density emerges from Josephson's relation as an interplay
between the condensate density in a finite size system, and the infrared
structure of the single particle Green's function.  We then discuss the
trapped two-dimensional Bose gas, where the interaction changes the transition
qualitatively from Bose-Einstein in an ideal gas to a KTB transition in the
thermodynamic limit.  We show that the transition temperature lies below the
ideal Bose-Einstein transition temperature, and calculate the first correction
in terms of the interparticle interactions.  The jump of the total superfluid
mass at the transition is suppressed in a trapped system.

\end{abstract}


\maketitle

\section{Introduction}

    The ability to produce a two-dimensional atomic gas trapped in an optical
potential has stimulated interest in the Kosterlitz-Thouless-Berezinskii (KTB)
transition in such a system \cite{Ketterle,2Dexp,smerzi,simula}.  Recently,
interesting phase patterns have been measured in a quasi two-dimensional
situation \cite{dalibard}.  A homogeneous Bose gas in two dimensions undergoes
Bose-Einstein condensation only at zero temperature, since long wavelength
phase fluctuactions destroy long range order \cite{Hohenberg}; nonetheless
interparticle interactions drive a phase transition to a superfluid state at
finite temperature, as first pointed out by Berezinskii \cite{Berezinskii} and
by Kosterlitz and Thouless \cite{KT}.  The phase transtion is characterized by
an algebraic decay of the off-diagonal one-body density matrix below the
transition temperature, $T_{KT}$.  Furthermore, the superfluid mass density,
$\rho_s$, jumps with falling temperature, $T$, from 0 just above $T_{KT}$ to a
universal value $\rho_s=2m^2 T_{KT}/\pi$ just below \cite{Nelson}, where $m$
is the atomic mass.  We use units $\hbar=k_B=1$.

    In a non-interacting homogeneous Bose gas in two dimensions, the density
is given in terms of the chemical potential $\mu$ by
\beq
   n\lambda^2 = -\log[1-e^{\beta \mu}],
\label{nc}
\eeq
where $\beta =1/T$, and $\lambda=(2 \pi/mT)^{1/2}$ is the thermal
wavelength.  As $\mu$ approaches zero, the phase-space density grows
arbitrarily, implying that the ideal Bose gas does not undergo Bose
condensation at finite temperature.  On the other hand, a trapped
non-interacting gas undergoes a Bose-Einstein condensation at a finite
temperature \cite{kleppner}.  In the semi-classical limit, the total number of
particles is given by $N(\beta\mu) = g_2(-\beta\mu)(T /\omega)^2$, where
$g_2(x)=\sum_{n=1}^\infty e^{-nx}/n^2 \simeq \pi^2/6-x(1-\log x)$, for $x \to
0$, and $\omega$ is the trapping frequency; thus the condensation temperature
is $T_{BEC}=\sqrt{6N}\omega/\pi$.

    The question of whether an interacting trapped Bose gas undergoes an
ordinary Bose-Einstein or a KTB has not completely been settled \cite{dima}.
In this paper we show that in the thermodynamic limit -- in which the total
particle number $N$ goes to $\infty$, and the trap frequency, $\omega$, goes
$0$, with $N \omega^2$ constant -- interactions at the mean field level
destroy ordinary Bose condensation; instead, the system undergoes a KTB
transition at a temperature slightly below the ideal condensation temperature,
$T_{BEC}$.

    We approach the phase transition of the homogeneous two-dimensional Bose
gas by carrying out a scaling analysis (Sec.  II and III) similar to the one
we used earlier to discuss the phase transition in a dilute three-dimensional
homogeneous Bose gas \cite{club,ours,others}.  The phase below $T_{KT}$ is
characterized by an algebraic decay in space of the single particle Green's
function, $G$.  As in three dimensions, the transition point occurs when the
single particle spectrum becomes gapless, $G^{-1}(0,0)=0$, where
$G(k,z_{\nu})$ is the Fourier transform of $G$ in space and time.  We first
discuss, in Sec.  II, the scaling structure of $G$ as $T_{KT}$ is approached
from above, and rederive the relation between the temperature and density at
the transition:
\beq
 \frac{T_{KT}}{n}= \frac{2\pi}{m \log(C/\alpha)},
 \label{tnc}
\eeq
where $C$ is a constant \cite{pierre}, $\simeq (380 \pm 3)/2 \pi$
according to numerical simulation \cite{Svistunov2D}, and we take the
interparticle coupling constant in two dimensions to have the form $g=2 \pi
\alpha/m$, where $\alpha$ is a dimensionless parameter \cite{scatter}.

    We then turn, in Sec.  III, to the structure below $T_{KT}$, working in
terms of the Green's function in momentum space.  Below $T_{KT}$ we expect
$G(r-r')\sim 1/|r-r'|^{\eta}$ for an infinite system and $|r-r'|\to \infty$,
where $\eta$ depends on $T$ (and equals 1/4 at $T_{KT}$ \cite{KT}); however,
$G(k)$ is not well-defined in an infinite system.  We therefore analyze the
structure starting with a finite size system, of characteristic dimension $L$;
such a system has a non-zero condensate density, $n_0 \sim G(r\sim L) \sim
1/L^{\eta}$, which vanishes in the thermodynamic limit, $L\to\infty$.
Nonetheless, we show using Josephson's relation between the condensate
density, the superfluid mass density, $\rho_s$, and the infrared behavior of
$G(k)$ \cite{josephson,standrews,jose1},
\beq
  \rho_s = -\lim_{k \to 0} \frac{n_0 m^2 }{k^2 G(k,0)},
  \label{josephson0}
\eeq
that $\rho_s$ is finite in the thermodynamic limit.  Our analysis of the
scaling structure is valid for all coupling strengths, and thus, going beyond
perturbation theory, is an extension of the work of Popov \cite{popov}.

    Then, in Sec.  IV, we apply our results to a trapped gas within the
local density approximation.  We show that the KTB transition
temperature, for a weakly interacting system, lies below the Bose-Einstein
condensation temperature of the ideal trapped gas by terms of order
$\alpha\log^2\alpha$.  Furthermore, the jump of the total superfluid mass at
the transition is $\sim \alpha$, and is thus highly suppressed compared with
that in a homogeneous system.

\section{Scaling structure above the transition}

    In this section we derive the KTB transition of a two-dimensional weakly
interactiong homogeneous Bose gas by studying, as in \cite{club}, the scaling
structure of $G$ just above the transition.  For wavevector $k$ and complex
frequency $z$, $G(k,z)$ is given in terms of the self-energy, $\Sigma(k,z)$,
by
\begin{equation}
   G^{-1}(k,z)=z+\mu - \frac{k^2}{2m} -\Sigma(k,z),
\end{equation}
where $\mu$ is the chemical potential.  The transition is defined, for
fixed density, by the temperature at which the single particle spectrum
becomes gapless, and consequently off-diagonal single particle density matrix
decays algebraically; at the transition point, $G^{-1}(0,0)=0$, or
$\mu=\Sigma(0,0)$.  In terms of $G$ the density is
\begin{equation}
  n=-T \sum_\nu \int \frac{d^2k}{(2 \pi)^2} G(k,z_\nu).
\end{equation}
where the $z_\nu=2 \pi i\nu T$ are the Matsubara frequencies ($\nu=0,\pm
1, \pm 2, \dots)$.

    The mean-field contribution, $\Sigma_{mf}=2g n$, to the self energy, is
independent of frequency and momentum, and can be absorbed in a shift of the
chemical potential.  We introduce the mean-field coherence length, $\zeta$,
by
\beq
\frac{1}{2m \zeta^2}= \Sigma_{mf}-\mu;
  \label{zeta}
\eeq
at the transition,
\beq
  1/2m\zeta^2=\Sigma_{mf}- \Sigma(0,0).
 \label{zetatrans}
\eeq
In terms of $\zeta$, the mean-field Green's function is,
\beq
   G_{\rm mf}(k,z_\nu) =
  -\frac{2m\zeta^2}{(k\zeta)^2+1-8i\pi^2(\zeta/\lambda)^2\nu}.
 \label{Gmf}
\eeq
In the zero Matsubara frequency sector, the self-energy has the scaling
structure:
\begin{equation}
  \Sigma(k,0)-\Sigma_{mf}= \frac{1}{2m}
  \sigma_0\left(k \zeta, \frac{\alpha \zeta^2}{\lambda^2} \right),
 \label{sig0}
\end{equation}
where $\sigma_0$ is dimensionless and $\sim \alpha^2$ as $\alpha \to 0$.
In contrast to the case of three-dimensions, self-energy diagrams beyond
mean field are ultraviolet convergent even in the zero Matsubara frequency
sector.

    Using Eqs.~(\ref{sig0}) and (\ref{zeta}) we may write the transition
condition, (\ref{zetatrans}),  as
\beq
  \sigma_0\left(0, \frac{\alpha \zeta^2}{\lambda^2} \right) + 1 =0;
\eeq
at the phase transition the parameter $\alpha \zeta^2/\lambda^2 \equiv J$
approaches a finite value(fixed point), $J^*$, determinined by this equation.
The contribution of non-zero Matsubara frequencies in the denominator of
Eq.~(\ref{Gmf}) is $\sim (\alpha \zeta^2/\lambda^2)/\alpha$, and thus in the
limit of small $\alpha$, the contributions of non-zero Matsubara frequencies
to the self-energy are of relative order $\alpha$ and higher, and can be
neglected.

    To calculate the critical density at given temperature we use the
mean-field density as reference:
\begin{eqnarray}
   n_c(\alpha,T) & = & n_{\rm mf}(\mu = -1/2m\zeta^2)
  \nonumber \\
     &-& T\sum_\nu \int \frac{d^2k}{(2\pi)^2}\left[G(k,z_\nu)- G_{\rm
  mf}(k,z_\nu)\right].
\nonumber \\
 \label{NC0}
\end{eqnarray}
In leading order, we neglect non-zero Matsubara frequencies in the summation
on the right, and derive
\begin{eqnarray}
   \lambda^2 n_c = \log \frac{4 \pi \zeta^2}{\lambda^2}
  &+& 2 \int k\, dk
\left(\frac{1}{k^2+\zeta^{-2}(1+\sigma_0)}\right.\nonumber\\
  &&-\left. \frac{1}{k^2+\zeta^{-2}}\right) +{\cal O}(\alpha),
\label{TC2}
\end{eqnarray}
Using Eq.~(\ref{sig0}) we see that the integral in (\ref{TC2}) is a
constant of order unity, independent of $\alpha$.  Thus we arrive at the
critical density,
\begin{equation}
  \lambda^2 n_c = \log \frac{C}{\alpha} +{\cal O}(\alpha),
  \label{nckt}
\end{equation}
as in Eq.~(\ref{tnc}).

    The true correlation length, $\xi$, above $T_{KT}$ is given by
\beq
    \frac1{2m\xi^2} \equiv \Sigma(0,0) - \mu = \Sigma(0,0)- \Sigma_{mf} +
  \frac1{2m\zeta^2}.
\eeq
To make contact with the theory of critical phenomena, we rewrite this
equation as
\beq
  M(J)= \frac{\lambda^2}{\alpha \xi^2} = \frac1J +
  \frac{2m\lambda^2}{\alpha} \left( \Sigma(0,0) - \Sigma_{mf} \right),
\eeq
where from the previous discussion we know that the right side is a
function only of $J =\alpha \zeta^2/\lambda^2$.  At the transition, $J$ goes
to the fixed point, $J^*$, given by $M(J^*)=0$.  Furthermore, approaching the
transition from above, we see from Eq.~(\ref{TC2}) that $\Delta J= J - J^*
\propto (T - T_{KT})/\alpha$.  The dependence of $M(J)$ on $\Delta J$ near the
fixed point determines the critical index $\nu$, of the correlation length,
$\xi \sim |T-T_{KT}|^{-\nu}$, in particular that $M(J) \sim |\Delta J|^{2\nu}
\sim |T-T_{KT}|^{2\nu}$, in the neighborhood of the fixed point.

\section{Scaling structure below the transition}

    At the transition temperature, $T_{TK}$, the single particle Green's
function decays algebraically.  Below, the scaling structure is most readily
analyzed in momentum space, as above; however this approach is made difficult
by the fact that below $T_{TK}$ the single particle Green's functions
continues to decay algebraically in real space sufficiently slowly that its
Fourier transform is not absolutely convergent.  In order to avoid this
problem, and to use the same approach as above, we adopt the strategy of
working in a finite size system, of characteristic dimension $L$, in which the
condensate density, $n_0$, is non-zero.  At the very end, we take the
thermodynamic limit, $L\to\infty$, at fixed density, in which case $n_0$ also
goes to zero, since long-range order is prohibited in two dimensions.

    In a finite size system the condensate density is given by
$n_0=n-\tilde{n}$, where
\beq
   \tilde{n} = -T \sum_n \int_{k_0}^{\infty} \frac{d^2k}{(2\pi)^2}G(k,n),
\eeq
and $k_0 \sim \pi/L$; in leading order, $G$ here can be taken to be the
infinite size Green's function.  Since for $T\le T_{KT}$,
\beq
    G(k\to 0,0)= -2mK \zeta^\eta/k^{2-\eta},
  \label{Gk0}
\eeq
where the constants $K$ and $\eta$ ($0\le\eta\le 2$) in general depend on $T$,
we find to leading order,
\begin{equation}
   n_0^L = \frac{2K}{\eta\lambda^2} (k_0 \zeta)^{\eta},
  \label{n0L}
\end{equation}
The result (\ref{n0L}) is given for a circular box; in general it is modified
by a numerical factor close to unity, weakly dependent on the geometry.

    Below $T_{KT}$ the system is superfluid even though the condensate density
is not extensive and vanishes in the thermodynamic limit.  However, in two
dimensions, an algebraically decaying correlation function is sufficient to
yield a non-vanishing superfluid density, $\rho_s$, in the thermodynamic
limit, as can be seen from the Josephson relation
\cite{josephson,standrews,jose1,jose} between $\rho_s$ and $n_0$ in a finite
system:
\begin{eqnarray}
  \rho_s = -\lim_{k \to k_0} \frac{n_0 m^2 }{k^2 G(k,0)}.
  \label{josephson}
\end{eqnarray}
Thus from Eqs.~(\ref{Gk0}) and (\ref{n0L}) and find
\begin{equation}
   \rho_s=\frac{m^2 T}{2 \pi \eta}.
 \label{eta}
\end{equation}
Inverting the relation we obtain $\eta=m^2 T/2 \pi \rho_s$; thus this
prediction of the spin-wave approximation for the critical exponent is
prediction below $T_{KT}$.  Since $\rho_s \le mn$, we see that
\beq
   \eta \ge \frac1{n\lambda^2}.
\eeq
Furthermore, since $\eta \le 2$, the superfluid mass density can never
vanish at finite temperature below $T_{KT}$, and thus it must be discontinuous
across the phase transition.  At $T_{KT}$, the value of the universal jump of
the superfluid density predicted by Nelson and Kosterlitz \cite{Nelson},
$\rho_s = 4T_{KT}m^2/2\pi$, leads to $\eta=1/4$.

    Let us now discuss the detailed structure of the Green's function below
the transition.  We basically follow the scaling approach used in \cite{N0} in
three dimensions.  Our strategy is to expand the self-energies formally in
powers of $\alpha$, $n_0$, and $k_0$.  In the infinite size system, $k_0\to
0$, the self-energies diverge as $n_0\to 0$; the point, $k_0\to 0$, $n_0\to 0$
is singular.  However Josephson's relation constrains the limit $n_0\to 0$ and
$k_0\to 0$ in terms of $\rho_s$.

    The particle density, $n$, in the condensed phase is a function of
$\alpha$, $n_0$, and $T$, and has the form, $n(\alpha,n_0,k_0,T)=n_0 +
\tilde{n}(\alpha,n_0,k_0,T)$, where $\tilde{n}(\alpha,n_0,T)$ is the density of
non-condensed particles (with momentum $k >k_0$).  At the transition
temperature, $\tilde{n}(\alpha,0,T_c)=n_c$.  We calculate $\tilde{n}(a,n_0,T)$
in terms of the matrix Green's function,
\beq
  {\cal{G}}(rt,r't')= -i \left(\langle
  T\left(\Psi(rt)\Psi^\dagger(r't')\right)\rangle\right. \nonumber\\
  - \langle
  \Psi^\dagger(r't')\rangle \langle \Psi(rt)\rangle\left.\right),
\eeq
where the two component field operator is $\Psi(rt)=
\left(\psi(rt),\psi^\dagger(rt) \right)$.  The Fourier components of
${\cal{G}}^{-1}$ have the form,
\beq
  {\cal{G}}^{-1}(k,z_n) =
   \pmatrix{
  z_n +\mu-\varepsilon_k -\Sigma_{11} & \hspace{-18pt} -\Sigma_{12}\cr
  -\Sigma_{21}&\hspace{-18pt} -z_n + \mu-\varepsilon_k -\Sigma_{22} \cr }
\nonumber\\
\eeq
where $\varepsilon_k=k^2/2m$, and the $\Sigma_{ij}(k,z_n)$ are the
corresponding self-energies.  The chemical potential, $\mu$, depends on $n_0$
and $k_0$, and is specified by the Hugenholtz-Pines relation \cite{HP},
\beq
  \mu = \Sigma_{11}(0,0)-\Sigma_{12}(0,0).
  \label{mu}
\eeq

    The lowest order mean field self-energies, $\Sigma_{11}=\Sigma_{11}^{mf}
=2 g (n_0 + \tilde{n})$, $\Sigma_{12}=\Sigma_{12}^{mf}=g n_0$ are independent
of momenta and Matsubara frequency, and, as above $T_{KT}$, we absorb them in
a mean field coherence length, $\bar{\zeta}$,
\begin{eqnarray}
 \mu-(\Sigma_{11}^{mf}&-&\Sigma_{12}^{mf})\nonumber\\
  &=&(\Sigma_{11}(0,0)-\Sigma_{12}(0,0))-
   (\Sigma_{11}^{mf}-\Sigma_{12}^{mf}) \nonumber\\ &\equiv& -1/2m
  \bar{\zeta}^2.
\label{HPself}
\end{eqnarray}

    Since the propagators remain formally infrared convergent we can derive
the scaling structure of the self-energies by power-counting.  As above the
transition, we may neglect non-zero Matsubara contributions to leading order.
As opposed to the three-dimensional case, there are no formal ultraviolet
divergencies in the expansion beyond those in mean field, and therefore need
for renormalization.  The expansion of the self-energies beyond mean field
starts at order $\alpha^2 \bar{\zeta}^2/\lambda^2$; furthermore $\Sigma_{12}$
is formally at least of order $n_0$.  Diagrams of order $g^\kappa$ with
$\kappa \ge 3$ in the formal expansion contain vertices with two Green's
functions entering; similar to the structure at $T_c$, they involve the
dimensionless combinations $\alpha \bar{\zeta}^2/\lambda^2\equiv P$ and $n_0
\lambda^2$.  The latter part originates from the dependence of ${\cal G}^{mf}$
on $2 m\Sigma_{12}^{mf}\sim \alpha n_0$.  Any diagram with an explicit power,
$p$, of $n_0$ can be generated from a corresponding diagram of power $p-1$ in
which a line is replaced by $\sqrt n_0$ at each of its ends.  Thus each power
of $n_0$ involves one fewer two-momentum loop to be integrated over.  The
explicit $n_0$ dependences enter in two ways.  Terms involving $G_{11}-G_{12}$
lead to the combination $P^2 n_0 \lambda^2$, which vanishes as $n_0\to 0$.  On
the other hand terms involving the combination $G_{11}+G_{12}$, which in mean
field diverges as $n_0\to 0$ in the infrared limit, lead to divergences which
are cutoff by $k_0$ and thus produce an additional
$(k_0\bar\zeta)^2/(n_0\lambda^2P)$ dependence.  In the limit $n_0\to 0$,
$k_0\to 0$, only the dependence on $(k_0\bar\zeta)^2/n_0\lambda^2\equiv Q$
survives.

    Then with all momenta $k$ scaled by $1/\bar{\zeta}$, we find the following
scaling structure for the self-energies in this limit,
\begin{eqnarray}
   2m\bar\zeta^2\left(\Sigma_{ij}(k)-\Sigma_{ij}^{mf}(0)\right)
   &=&  \sigma_{ij}(k\bar{\zeta},P, Q),
 \label{sigma}
\end{eqnarray}
where the $\sigma_{ij}$ are dimensionless functions of dimensionless
variables, and we neglect terms proportional to positive powers of $n_0$ and
$k_0$.  In particular, for $k\to 0$,
\begin{eqnarray}
  2m\bar\zeta^2\left((\Sigma_{11}(0,0)-\Sigma_{12}(0,0))
   -(\Sigma_{11}^{mf}\Sigma_{12}^{mf})\right)&& \nonumber\\
   =s(P,Q),
\label{self2}
\end{eqnarray}
where $s$ is a dimensionless function.  Comparing with Eq.~(\ref{HPself})
we see that $s+1=0$, an equation that determines $P$ as a
function of $Q$, so that
\begin{equation}
    \bar{\zeta} = \frac{\lambda}{\sqrt{\alpha}}h(Q),
  \label{zeta1}
\end{equation}
where $h$ is dimensionless.

    We now take the limit $n_0\to0$ and $k_0\to 0$.  Equations~(\ref{HPself})
-(\ref{self2}) imply that $G_{11}(k_0,0) = -m\bar\zeta^2\ell(Q)$, and thus the
Josephson relation implies that the superfluid mass density has the structure,
\beq
   \rho_s = \frac{m}{\lambda^2 Q\ell(Q)},
\eeq
where $\ell$ is a dimensionless function; this equation relates $n_0$ and
$k_0$ to $\rho_s$ in the limiting process.  In this limit $P$ as well as $Q$
are constants dependent only on $\rho_s/T$.  Similarly the other dimensionless
functions depend in this limit only on $\rho_s$.  These forms are valid over
the temperature range from $T_{KT}$ down to $T=0$ (with non-zero Matsubara
frequency contributions neglected); this entire region is therefore critical.

    We can gain further insight into the structure below $T_{KT}$ by
reformulating it in terms of the correlation lengths, $\zeta_T$ and $\zeta_L$,
that control the infrared behavior of the transverse and longitudinal Green's
functions, $G_T=G_{11}-G_{12}$, and $G_L=G_{11}+G_{12}$.  We regularize the
infrared divergent structure below $T_{KT}$ by assuming a finite condensate
density, $n_0$, and finite correlation lengths.  In this description, the
low temperature phase of the KTB transition is characterized by the ratio of
amplitude (L) and phase (T) fluctuations of the order parameter, even in the
absence of long-range order in the thermodynamic limit in which $n_0 \to 0$.
In the end we take the limit, $n_0\to 0$, $\zeta_T\to \infty$, and $\zeta_L\to
\infty$.  In this way, we do not have to introduce an explicit infrared
cutoff, $k_0$, as we had to above.  We define
\beq
  \frac1{2m\zeta_T^2} = \Sigma_{11}(0,0)-\Sigma_{12}(0,0)-\mu,
   \label{zetaT}
\eeq
and
\beq
  \frac1{2 m \zeta_L^2} = \Sigma_{11}(0,0)+\Sigma_{12}(0,0)-\mu,
\label{zetaL}
\eeq
in terms of which $G_T$ and $G_L$ are given by,
\beq
  G_T^{-1}(k,0)=- \left( \frac{k^2}{2m}
  + [\Sigma_{11}(k,0)-\Sigma_{12}(k,0)] \right.\nonumber\\ \left.
  - [\Sigma_{11}(k,0) -\Sigma_{12}(0,0)]
  +\frac1{2m \zeta_T^2} \right).
 \eeq
and
\beq
  G_L^{-1}(k,0)=- \left( \frac{k^2}{2m}
  + [\Sigma_{11}(k,0)+\Sigma_{12}(k,0)] \right.\nonumber\\ \left.
  - [\Sigma_{11}(k,0) +\Sigma_{12}(0,0)]
  +\frac1{2m \zeta_L^2}\right).
\eeq

    The equilibrium state of the system is specified below the transition by
$\zeta_T \to \infty$, as seen from the Hugenholtz-Pines relation (\ref{mu}),
as well as $n_0\to 0$.  From Eqs.~(\ref{zetaT}) and (\ref{zetaL}), we obtain
the ratio
\beq
 \left(\frac{\zeta_T}{\zeta_L}\right)^2=1 + 4 m\zeta_T^2 \Sigma_{12}(0,0).
\label{zetaTL}
\eeq
The second term is of order $n_0\zeta_T^2$.  From the previous discussion,
we deduce that in the limit $n_0 \lambda^2 \to 0$, $4m\zeta_T^2
\Sigma_{12}(0,0) \to 8 \pi \alpha n_0 \zeta_T^2 f(\zeta_T/\zeta_L)$, where $f$
is a dimensionless function.  Equation~(\ref{zetaTL}) thus implies that the
ratio $\zeta_T/\zeta_L$ is a function only of $n_0 \zeta_T^2$.  We therefore
we arrive at a self-consistent solution in the limit $n_0 \to 0$ and $\zeta_T
\to \infty$ in which $n_0\zeta_T^2$ remains finite.  The value of this
parameter is determined in terms of $\rho_s$ by Josephson's relation, which we
write in the form
\cite{jose1}
\beq
   \rho_s=\lim_{\zeta_T \to \infty}
   m n_0 \Big( 1+  \hspace{130pt}\nonumber\\
   2m\frac{\partial}{\partial k_z^2}
   [\Sigma_{11}(k,0)-\Sigma_{12}(k,0)] \big|_{k=0} \Big).
\eeq
For finite $\zeta_T$, the derivative of the self energies with respect to
$k^2$ are finite in the limit $k \to 0$, but must diverge as $\zeta_T^2$.
The right side is a function only of $n_0 \zeta_T^2$ in the limit
$\zeta_T^2\to \infty$, and $n_0\to 0$, thus defining $n_0 \zeta_T^2$ in terms
of $\rho_s$.

\section{The transition in a trapped gas}

    We turn now to the behavior of a two-dimensional system trapped in an
oscillator potential, of frequency $\omega$.  We consider for simplicity only
the thermodynamic limit $N \to \infty$, $\omega \to 0$, with $N \omega^2$
constant, where $N$ is the total particle number.

    In the absence of interactions the system undergoes a Bose-Einstein
condensation at the critical temperature $T_{BEC}=\sqrt{6N}\omega/\pi$.
However, at the mean field level, interactions destroy simple Bose-Einstein
condensation.  To see how the critical temperature is reduced to zero, we
begin with the density profile calculated in the local density approximation,
\begin{eqnarray}
  n_{\rm mf}(r) & = &
   \int \frac{d^2k}{(2\pi)^2}
 \frac{1}{e^{\beta (k^2/2m +V_{\rm eff}(r) - \mu)}-1}
   \nonumber\\
 &=& -\frac1{\lambda^2}
 \log \left( 1- e^{\beta (\mu-V_{\rm eff}(r))}\right),
 \label{nmf}
\end{eqnarray}
where $V_{\rm eff}(r)= m \omega^2 r^2/2 + 2gn(r)$.  Since $n(r)$ has negative
curvature at the origin, interactions tend to reduce the effective trapping
frequency.  Explicitly, expanding (\ref{nmf}) to order $r^2$ about the origin,
we find that the self-consistent trap frequency at the origin is given by,
\begin{eqnarray}
   m\omega_{\rm eff}^2 &\equiv& m\omega^2 +2g\left(\frac{\partial^2
      n_{\rm mf}(r)}{\partial r^2}\right)_{r=0}
\nonumber\\
  &=& m\omega^2 \frac{1-e^{\beta(\mu -2gn(0))}}
       {4\beta g\lambda^{-2} + 1-e^{\beta(\mu -2gn(0))}}.
\end{eqnarray}
As $\mu\to 2gn(0)$, the limit in which Bose-Einstein condensation occurs
in mean field, we have
\beq
  \omega_{\rm eff}^2 \simeq \omega^2 \frac{\beta}{2\alpha}\left(2gn(0)-
   \mu\right);
\eeq
the potential becomes arbitrarily flat as $\mu\to 2gn(0)$.  In the
thermodynamic limit, $N\omega^2$ constant, the effective oscillator length
$d_{\rm eff} \equiv 1/\sqrt{m \omega_{\rm eff}}$ is $\sim \left(NT/(2gn(0)-
\mu)\right)^{1/4}$.  At the point of mean-field Bose condensation, $2gn(0)-\mu
\equiv 1/2m\zeta(0)^2$ is therefore $\sim T/N$; using the estimate (\ref{nc}),
we see that the number of particles in the center, $\sim n(0)\pi d_{\rm
eff}^2$, grows as
$(T/\omega)\sqrt{\alpha}(\zeta(0)/\lambda)\log(\zeta(0)/\lambda)$.  Now, an
extensive population of the condensate mode implies $\zeta(0)/\lambda \sim
N^{1/2}$ within mean field.  From our estimate of the number of (excited)
particles in the center of the trap we obtain an upper bound for the critical
temperature of Bose condensation, $\sim \omega N^{1/2}/(\alpha^{1/2} \log N)$.
In the thermodynamic limit of the trapped system, the critical temperature of
Bose condensation goes to zero.

    Mean fields destroy Bose-Einstein condensation of an ideal gas.  However,
the system does undergo a KTB transition in the thermodynamic limit, at a
temprature calculable, to leading order in $\alpha$, in terms of the mean
field density profile, Eq.~(\ref{nmf}).  The KTB transition occurs when the
chemical potential reaches the critical chemical potential, $\Sigma(0,0)$,
calculated for a homogeneous system of the same density and temperature as in
the center of the trap, or equivalently, Eq.~(\ref{nckt}) is satisfied.  As we
see from Eqs.~(\ref{NC0}) and (\ref{TC2}) in the homogeneous case, the
critical density is given to logarithmic accuracy by the mean field density
evaluated at the critical $\zeta$ given by Eq.~(\ref{zetatrans}); the
corrections arise from critical fluctuations, which are, however, important
only at small distances where $\frac12m\omega^2r^2\simle |gn(0)-\mu|$, or
\beq
    r \simle r_c \simeq \frac{1}{m\omega \zeta}.
 \label{rc}
\eeq
In the thermodynamic limit the critical region produces corrections to the
total number, $\Delta N/N\sim \alpha^2\log\alpha$, which can be neglected, and
thus, to leading order in $\alpha$, we can calculate the transition from
density profile, Eq.~(\ref{nmf}).

    Expanding $n_{\rm mf}(r)$ to first order in $\alpha$ we have,
\beq
   n_{\rm mf}(r) = n_0(r)
       + \frac{2\beta gn_0(r)}{\lambda^2} \frac{1}{e^{\beta m\omega^2r^2/2
  -\beta\mu}-1}
\eeq
where
\beq
  n_0(r) =  -\frac1{\lambda^2}
      \log \left( 1- e^{\beta (\mu- m\omega^2 r^2/2)}\right),
\eeq
is the ideal gas density.  Integrating over space, we find the first
correction to the total number of particles,
\beq
   N &=& \int d^2r n_{\rm mf}(r) =\int d^2r n_0(r)  - \alpha
            \left(\frac{2\pi n_0(0)}{m\omega}\right)^2
 \nonumber\\
     &=& g_2(-\beta\mu)\frac{T^2}{\omega^2} - \alpha
            \left(\frac{2\pi\log(-\beta\mu)}{m\omega\lambda^2}\right)^2,
\eeq
where $\mu = 2gn(0)- 1/2m\zeta^2$.  At the Kosterlitz-Thouless
transition, $\mu \sim \alpha$.  To within logarithmic corrections we may
replace $n_0(0)$ by $-\lambda^{-2}\log(\alpha)$, and $g_2(-\beta\mu)$ by
$g_2(0)$.  Thus at given $T$,
\beq
  N_{KT}(T,\alpha) = N_{\rm BEC}(T) - \alpha
            \left(\frac{2\pi\log\alpha}{m\omega\lambda^2}\right)^2,
\eeq
where $N_{\rm BEC}(T) = (\pi^2/6)(T/\omega)^2$ is the relation between
particle number and temperature of the ideal Bose gas transition in the
trap.  Equivalently
\beq
  \frac{T_{KT} - T_{BEC}}{T_{BEC}} = - \frac{3}{\pi^2}\alpha \log^2\alpha
   +{\cal O}(\alpha\log\alpha).
\eeq
The existence of the Kosterlitz-Thouless transition just below $T_{\rm
BEC}$ in a trap involves physics beyond mean-field.

    In a KTB transition in a homogeneous system, the superfluid mass density
jumps with falling temperature discontinuously at $T_{KT}$ from zero to
$2m^2T_{KT}/\pi$.  In a trap, however, the transition first occurs in the
center, and extends over a region of size $r_c$, Eq.~(\ref{rc}).  The total
superfluid mass, $M_s$, is therefore
\beq
   M_s = \int \rho_s(r) d^2 r \sim \pi \rho_s(r=0) r_c^2,
\eeq
so that
\beq
    \frac{M_s}{M} \sim \frac1N \left(\frac{T_c}{\omega}\right)^2
      \frac{\lambda^2}{\zeta^2} \sim \frac{\alpha}{P},
\eeq
where $M=mN$, and $P = \alpha\zeta^2/\lambda^2$ is a constant (fixed
point) at the central transition point.  The jump in $M_s$ is thus highly
suppressed, and goes to zero in the limit of an ideal gas.

    A key indication of superfluid behavior below the transition would be
creation of a vortex at the center of the trap, where the system first becomes
superfluid, e.g., by cooling a rotating system through $T_{KT}$.  In order to
create a vortex at $T_{KT}$ it is necessary that the vortex core, of radius
$\sim\zeta$, fit within the critical region, of size $\sim r_c$, at the
transition.  From Eq.~(\ref{rc}), $\zeta/r_c \sim m\omega\zeta^2$, so
that in the thermodynamic limit, $\zeta/r_c \sim P/\alpha\sqrt N$.  Thus
for large $N$, creation of a vortex within the critical region becomes
possible.  The critical rotation frequency, $\Omega_c$, for creation of a
vortex is of order $(1/mr_c^2)\log(r_c/\zeta)$, and therefore
$\Omega_c/\omega \sim (\zeta/r_c)\log(r_c/\zeta) \sim (1/\alpha\sqrt
N)\log(\alpha\sqrt N)$.

    A further probe of the state of the system below $T_{KT}$ would be
determination of the density correlations, as have been recently measured in
Bose gases trapped in optical lattices \cite{bloch}.  These correlations will
depend strongly on the amplitude fluctuations, described by $\zeta_T$
\cite{density}.

    This research was supported in part by NSF Grants PHY0355014 and
PHY0500914.  LKB and LPTMC are Unit\'es Associ\'ees au CNRS, UMR 8552 and UMR
7600.  Authors MH and GB are grateful to the Aspen Center for Physics where
this work was completed.


\begin{thebibliography}{99}
\bibitem{scatter} The scattering length in two dimensions is not a
well-defined concept since the scattering cross section vanishes in the limit
of low energies and low momentum.  Nonetheless, for a singular potential such
as a hard-core, it is sufficient to sum particle-particle scattering processes
in the T-matrix and to replace the bare potential by the T-matrix in the
Hamiltonian \cite{Schick}.  Since the T-matrix depends only logarithmically on
energy and momentum, we can still work to leading order with a momentum and
energy independent coupling constant, $g =2 pi/(m log 1/na^2)$, where $a$
corresponds to the hard core diameter.  We further neglect the density
dependence of the coupling constant in the following (it does not enter in an
essential way) and write $g=2 \pi \alpha/m$.
\end{thebibliography}
\end{document}